%% file: main.tex
\documentclass[sn-nature]{sn-jnl}
\usepackage{manyfoot}
\usepackage{ulem}
\usepackage{xspace}
\usepackage{graphicx}
\usepackage{fancyhdr}
\usepackage{pslatex}   
\usepackage{xcolor}
\usepackage{bm}
\usepackage{color}
\usepackage{amsmath, amsthm, amssymb} 
\usepackage{amssymb}
\usepackage{enumerate}
\usepackage{enumitem}
\usepackage[symbol]{footmisc}

\usepackage{hyperref}
\usepackage{multirow}
\usepackage{ulem}
\usepackage{comment}

\usepackage{lineno}
\usepackage{braket}
\usepackage[utf8]{inputenc}
\usepackage[T1]{fontenc}

\definecolor{darkviolet}{rgb}{0.58, 0.0, 0.83}

\usepackage{datetime}

\begin{document}


\title[]{Exploring baryon semi-leptonic decays through polarization and entanglement}

\author{\large\bf The BESIII Collaboration\footnote[2]{
A full list
of authors and affiliations appear in the online version of this paper.}}



\input{01_abstract}

\maketitle

\input{02_motivation}
\input{03_formalism}

\input{04_experiment}

\input{05_results}
\input{06_conclusion}

\bibliography{main}

\noindent

\input{07_methods}
\clearpage\section*{The BESIII Collaboration \label{app:collab}}\begin{sloppypar}\hyphenpenalty=5000\widowpenalty=500\clubpenalty=5000\input{authorlist_2025-05-08.tex}\end{sloppypar}
\clearpage\section*{Acknowledgement}\begin{sloppypar}\hyphenpenalty=5000\widowpenalty=500\clubpenalty=5000\input{acknowledgement_2025-05-08.tex}\end{sloppypar}

\end{document}

%% file: 01_abstract.tex
\abstract{\boldmath
The unitarity of the Cabibbo–Kobayashi–Maskawa (CKM) matrix is a cornerstone of the Standard Model (SM). 
Precise tests of this unitarity require independent determinations of its elements, such as $|V_{us}|$, which governs the transition between strange and up quarks.
Current measurements from kaon and tau decays show tensions that may hint at physics beyond the SM~\cite{Seng:2021nar,RBC:2018uyk,Hudspith:2017vew}.
Hyperon semi-leptonic decays provide alternative probes, but have remained largely untapped because previous experiments lacked sufficient kinematic information, rendering the measurements insensitive to the relevant form factors~\cite{ParticleDataGroup:2024cfk}.
Here we report measurements of the axial‑vector and weak‑magnetism couplings, as well as the first determinations of the absolute branching fraction and weak‑electricity coupling in $\Lambda \to p e^{-}\bar{\nu}_e$, achieved by exploiting the polarization and quantum entanglement of $\Lambda\bar{\Lambda}$ pairs produced at the $J/\psi$ resonance. 
Combining our results with recent lattice Quantum Chromodynamics calculations~\cite{Bacchio:2025auj} gives $|V_{us}|_{\rm LQCD}=0.2339\pm0.0041$, a model‑independent determination consistent with CKM unitarity. 
By pioneering the exploitation of polarization and quantum entanglement in baryon semi-leptonic decays, our method enhances single‑event sensitivity and it is broadly applicable to other baryon semi-leptonic decays, establishing the foundation for a systematic research program that can achieve precision comparable to that of kaon decays and provide a stringent independent test of the SM. }

%% file: 02_motivation.tex
\begin{figure}[htp]
  \begin{center}
\includegraphics[width=0.9\textwidth]{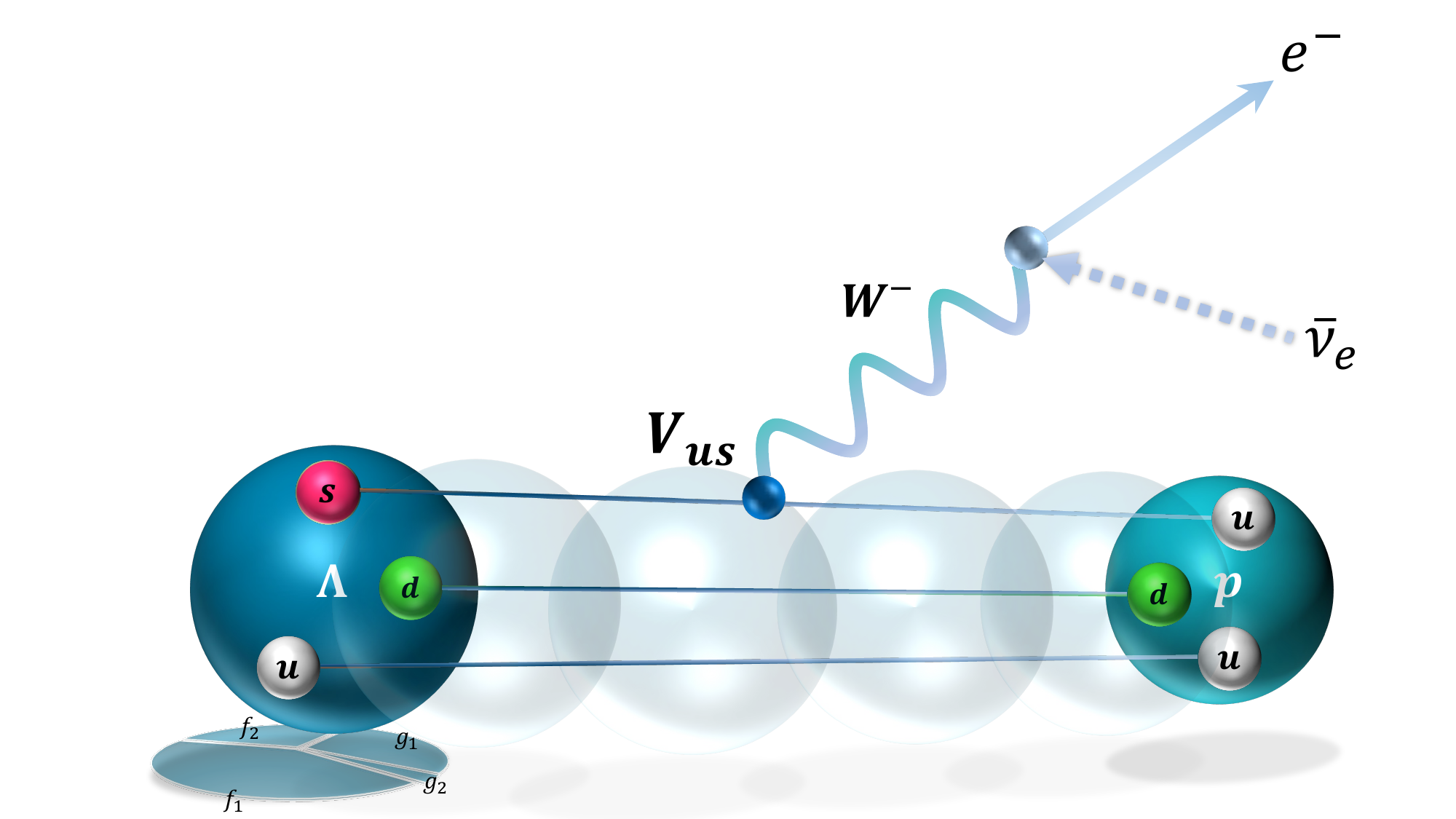}
      \caption{\textbf{Schematic representation of the $\bm \Lambda$ decay to a proton, an electron, and an electric anti-neutrino via a virtual (mediating) $\bm W^-$ boson. The pie chart shows the relevant magnitudes of the dominant vector and axial-vector form factors partaking in the transition. 
     }}
      
      \label{fig:formfactorsVus}
    \end{center}
\end{figure}

Transition form factors are fundamental hadron properties describing the dynamic behaviour of the transition between two states. The transition matrix element for a semi-leptonic weak decay of a hyperon is parameterized by six form factors, $f_i(q^2)$ and $g_i(q^2)$ ($i=1,2,3$). These represent the vector and axial vector parts of the weak interaction, respectively, and depend on the squared momentum transfer, $q^2$, of the intermediate $W$ boson. 
The vector form factors are commonly denoted as the vector $f_1(q^2)$, weak magnetism $f_2(q^2)$, and scalar $f_3(q^2)$ form factors. The axial-vector form factors are further subdivided into the axial $g_1(q^2)$, weak electricity $g_2(q^2)$, and pseudoscalar $g_3(q^2)$ form factors~\cite{Garcia:1985xz}. Each form factor is associated with a specific current and encodes information about the underlying hadronic structure.
Since both the scalar and pseudoscalar form factors are suppressed by the squared ratio of the lepton to hyperon mass, the electron--antineutrino decay modes effectively depend only on four form factors and $V_{us}$, as depicted in Figure \ref{fig:formfactorsVus}. 

When neglecting the $q^2$ dependence of the form factors, the branching ratio ${\cal B} $ of the semi-leptonic decay $\Lambda\to pe^-\bar{\nu}_e$ is given by~\cite{Garcia:1985xz}: 
\begin{equation}
\label{eq:Vusf1}
  {\cal B} 
    = \frac{\tau_\Lambda}{\hslash}G_F^2  |V_{us}|^{2} \frac{ \beta ^5 {M_\Lambda}^5}{60 \pi ^3}\left[ (1-\frac32\beta)(f_1^2+3g_1^2) - 4\beta g_1 g_2+ \frac27\beta^2\mathcal{F}_2 +{\mathcal O}(\beta^3)\right],
\end{equation}
 where $\beta=(M_{\Lambda}-M_{p})/M_{\Lambda}\approx 0.159$, with $M_{\Lambda}$ and $M_p$ denoting the masses of the $\Lambda$ and proton, respectively. The weak decay constant $G_F$ and the $\Lambda$ lifetime $\tau_{\Lambda}$ are known with high accuracy~\cite{ParticleDataGroup:2024cfk}. 
The form-factor-dependent function $\mathcal{F}_{2}$ is defined as: 
\begin{equation}
\begin{array}{ccll}
\mathcal{F}_2 = &3 f_1^2+3 {f_1} {f_2}+2 {f_2}^2+6 {g_1}^2+6g_2^2+21g_1g_2.\\
\end{array}\label{eq:F} 
\end{equation}
The ratios $g_1/f_1$\footnote{Throughout this Letter, $f_i\equiv f_i(q^2=0)$ and $g_i\equiv g_i(q^2=0)$ are implied
unless explicitly noted.}, $f_2/f_1$ and $g_2/f_1$ represent the axial-vector ($g_{av} \equiv g_1/f_1$), weak-magnetism ($g_w \equiv f_2/f_1$), and weak-electricity ($g_{av2} \equiv g_2/f_1$) couplings at zero momentum transfer $q^2$, respectively. Therefore, a precise measurement of the decay branching fraction can be used to determine $|V_{us}|$ if the form factors are known. The relative couplings $g_{av}$, $g_w$, and $g_{av2}$ can be determined from kinematic angular variables, but a reliable theoretical determination of $f_1$ requires an understanding of the subtle differences between the $d$- and $s$-quarks~\cite{Cabibbo:2003cu}. In this context, approximate flavor SU(3) symmetry serves as a good approximation, as confirmed by previous experimental results~\cite{Wise:1980iq, Bristol-Geneva-Heidelberg-Orsay-Rutherford-Strasbourg:1983jzt, Dworkin:1990dd}. 
This assumption treats the $s$ and $d$ quarks as equivalent, with symmetry breaking arising from their small mass difference relative to both $M_\Lambda$ and $M_p$.
For instance, $f_1$ is protected from leading-order SU(3)-breaking effects~\cite{Ademollo:1964sr}. However, corrections are essential at next-to-leading order. 
 There are various model estimates of $f_1$ using  quark models~\cite{Donoghue:1986th,Schlumpf:1994fb}, large-$N_c$~\cite{Flores-Mendieta:1998tfv, Flores-Mendieta:2004cyh, Flores-Mendieta:2024kvj, Flores-Mendieta:2025igu}, chiral expansions~\cite{Krause:1990xc,Anderson:1993as,Kaiser:2001yc}, and quantum chromodynamics (QCD) sum rules~\cite{Zhang:2024ick}. A drawback is that these models disagree on the size of the SU(3)-breaking corrections, which affects the precision of $|V_{us}|$~\cite{Mateu:2005wi}. During the last decade, the lattice QCD community has joined the efforts of determining  $f_1$ using a non-perturbative method~\cite{Guadagnoli:2005zs, Guadagnoli:2006gj, Bacchio:2025auj} and a model-independent approach~\cite{Guadagnoli:2004qw} developed for the form factors of meson decays~\cite{Hashimoto:1999yp, Becirevic:2004ya}.  Still, understanding the quark mass dependence of SU(3)-breaking remains a crucial issue. 
 
Another way to determine the vector form factor is indirectly via the $g_{av}$ coupling and $g_1$ form factor. The axial-vector coupling has been measured experimentally~\cite{Wise:1980iq, Bristol-Geneva-Heidelberg-Orsay-Rutherford-Strasbourg:1983jzt, Dworkin:1990dd}, with the most precise determination, $g_{av}=0.719\pm0.016\pm0.012$, performed more than thirty years ago by a Fermilab-based fix-target experiment using a neutral-hyperon beam~\cite{Dworkin:1990dd}, which observed $\sim37\cdot10^3$ $\Lambda\to pe^-\bar{\nu}_e$ candidates. The average value based on the Particle Data Group gives  $g_{av}=0.718\pm0.015$~\cite{ParticleDataGroup:2024cfk}. 
In contrast to $f_1$, the axial-vector form factor 
is not protected by the Ademollo-Gatto theorem~\cite{Ademollo:1964sr}. Thus, SU(3)-breaking corrections must be considered at leading order. During the last several years, the $g_1$ of hyperon beta decays have been calculated with high accuracy from first principles using the techniques of lattice QCD~\cite{Erkol:2009ev}. Although the hyperon axial form factor can be determined from the lattice, there is currently not enough experimental information.
The weak-electricity coupling, $g_{av2}$, is intimately linked to SU(3) symmetry. In the absence of second-class currents - which are characterized by their G-parity asymmetric transformation - it vanishes in the SU(3) limit~\cite{Weinberg:1958ut}. This quantity has so far only been determined once by the KTeV experiment for the process $\Xi^{0}\to \Sigma^{+} e^{-} \bar{\nu}_{e}$ decay~\cite{KTeV:2001djr} and was found to be consistent with zero. 

The CKM matrix element $|V_{us}|$  was evaluated by Cabibbo using the $\Lambda$ semi-leptonic decay~\cite{Cabibbo:2003ea,Cabibbo:2003cu}. The analyses were based on the determination of relative branching fractions~\cite{Wise:1980iq} and the $g_{av}$ coupling~\cite{Dworkin:1990dd} while the values of $f_2$ and $g_2$ were predicted within the conserved vector current (CVC) hypothesis~\cite{Gell-Mann:1960mvl} and SU(3) symmetry limit~\cite{Cabibbo:2003cu,Cabibbo:2003ea}, respectively. The obtained precision was found to be $\sigma(|V_{us}|)\sim0.0034$, where precise experimental measurements and methods are required to be competitive with the $V_{us}$ determined from kaon decays, $\sigma(|V_{us}|)\sim0.0009$~\cite{ParticleDataGroup:2024cfk}. Thus, in comparison to the traditional way of determining $|V_{us}|$ using kaon decays, where the accuracy is limited by knowledge of decay constants in lattice QCD calculation~\cite{FlavourLatticeAveragingGroupFLAG:2021npn}, more studies are required for the relevant transition matrix elements to reach a comparable level~\cite{FlavourLatticeAveragingGroupFLAG:2021npn}.

In this article, we determine the absolute branching fraction for $\Lambda\to pe^-\bar{\nu}_e$ and the couplings $g_{av}$, $g_{w}$, and $g_{av2}$ simultaneously for the first time. 
Meanwhile, two values of $|V_{us}|$ are provided, one under the assumption of SU(3) symmetry conservation, and the other using the lattice QCD determination of the form factors. Additionally, the product of $|V_{us}|$ and the dominant form factor combination, $|V_{us}|\cdot\sqrt{f_1^2+3g_1^2}$, has been extracted for the first time for any semi-leptonic baryon decay.

%% file: 03_formalism.tex
\begin{figure}[htp]
  \begin{center}
       \includegraphics[width=1\textwidth]{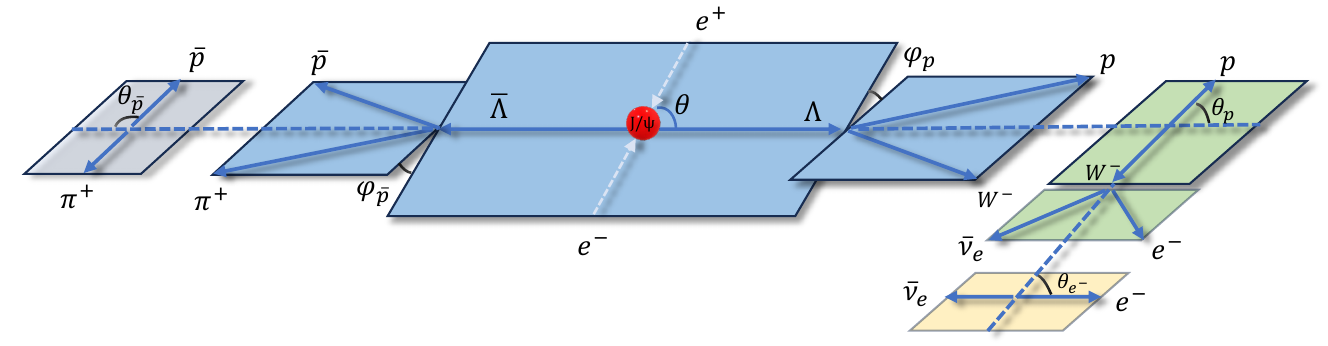}
      \caption{\textbf{Definition of the helicity angles for} $\bm{e^{+}e^{-}\to J/\psi \to \Lambda(\to pe^{-}\bar{\nu}_{e})\bar{\Lambda}(\to \bar{p}\pi^{+})}$. The angles $\theta$, $\theta_{\bar{p}}$, $\varphi_{\bar{p}}$, $\theta_{p}$, $\varphi_{p}$, and $\theta_e$ are the helicity angles of the $\Lambda$, $\bar{p}$, $p$, and lepton in the $e^+e^-$ c.m. system (blue plane), $\bar\Lambda$ rest frame (gray plane), $\Lambda$ rest frame (green plane), and $W^-$ rest frame (yellow plane), respectively.}    \label{fig:process}
    \end{center}
\end{figure}

The method, developed in Refs.~\cite{Perotti:2018wxm, Batozskaya:2023rek}, firstly exploits entangled polarized baryon–antibaryon pairs from the process
$e^+e^-\to J/\psi\to\Lambda(\to p e^- \bar{\nu}_e)\bar{\Lambda}(\to\bar{p}\pi^+)$ with its corresponding charge-conjugated $\Lambda\bar{\Lambda}$ decay modes. 
The production of the $\Lambda$ pairs with their subsequent decays are described with seven kinematic variables: six helicity angles and the four-momentum transfer between the $\Lambda$ and proton, $\boldsymbol{\xi} = \{\theta,\theta_p,\varphi_p,\theta_e,\theta_{\bar{p}},\varphi_{\bar{p}}, q^2\}$. 
The helicity angles are determined by boosting and rotating the particles into their respective helicity frames (Figure~\ref{fig:process}).
The production process is characterized by the $\Lambda$ scattering angle relative to the positron beam axis in the center-of-momentum (c.m.) frame, $\theta$. The angles $\theta_{p}$ and $\varphi_p$ ($\theta_{\bar{p}}$ and $\varphi_{\bar{p}}$) are defined with respect to the proton (anti-proton) direction in the reference system $\mathcal{R}_{\Lambda} (\mathcal{R}_{\bar{\Lambda}})$, where $\Lambda (\bar{\Lambda})$ is at rest and where the $\hat{z}$ axis points in the direction of the $\Lambda (\bar{\Lambda})$ in the c.m. system. The $\hat{y}$ axis is perpendicular to the production plane. 
The angle $\theta_e$ and the four-momentum transfer $q^2$ give the direction of the electron (positron) in the direction of the $\Lambda (\bar{\Lambda})$ in the $\mathcal{R}_{\Lambda}$ ($\mathcal{R}_{\bar{\Lambda}})$ frame.  
The structure of the seven-dimensional distribution of the kinematic variables is determined by six global parameters $\boldsymbol{\omega}=\{ \alpha_{\psi},\Delta\Phi,g_{av},g_{w},g_{av2},\alpha_+(\alpha_-) \}$ and can be expressed in a modular form as~\cite{Batozskaya:2023rek}:
\begin{equation}\label{eq:W}
    \mathcal{W}(\boldsymbol\xi;\boldsymbol\omega)=\sum_{\mu,\bar{\nu}=0}^3 C_{\mu\bar{\nu}}R_{\mu\kappa}(\Omega_p)b_{\kappa0}^{\Lambda}a_{\bar{\nu}0}^{\bar{\Lambda}} + c.c.,
\end{equation}
The $\mathcal{W}$ denotes the weights used in the maximum log-likelihood fit, as defined in Methods, for determining the form factors of interest. The production term $C_{\mu\bar{\nu}}(\theta;\alpha_{\psi},\Delta\Phi)$ is a $4\times4$ spin density matrix describing the polarization and the spin correlations between the $\Lambda$ pair and is defined in the reference frames $\mathcal{R}_{\Lambda}$ and $\mathcal{R}_{\bar{\Lambda}}$. The parameters $\alpha_{\psi}$ and $\Delta\Phi$ are related to two production amplitudes, where $\alpha_{\psi}$ governs the $\Lambda$ angular distribution and $\sin(\Delta\Phi)$ is proportional to the hyperon polarization~\cite{Perotti:2018wxm}. 
The $4\times4$ matrices $R_{\mu\kappa}b_{\kappa0}^{\Lambda}$ in Eq.~(\ref{eq:W}) represents the rotation and decay matrices between the $\Lambda$ and the proton in the semi-leptonic decay, respectively, while matrix $a_{\bar{\nu}0}^{\bar{\Lambda}}$ describes the transition from the $\bar{\Lambda}$ to the anti-proton in the non-leptonic decay via the asymmetry parameter $\alpha_+$ ($\alpha_-$ counts for $\Lambda\to p$ transition)~\cite{BESIII:2018cnd}. 
The elements of these matrices are parametrized in terms of the helicity angles as well as the weak decay parameters: $R_{\mu\kappa}(\theta_p,\varphi_p)b_{\kappa0}^{\Lambda}(\theta_e,q^2;g_{av},g_w,g_{av2})$ in reference system $\mathcal{R}_{\Lambda}$, $a_{\bar{\nu}0}^{\bar{\Lambda}}(\theta_{\bar{p}},\varphi_{\bar{p}};\alpha_+)$ in reference system $\mathcal{R}_{\bar{\Lambda}}$. The full expressions of $C_{\mu\bar{\nu}}$, $R_{\mu\kappa}b_{\kappa0}$ and $a_{\bar{\nu}0}$ are given in Refs.~\cite{Perotti:2018wxm,Batozskaya:2023rek}. Details of the coordinate system are provided in the Methods.

%% file: 04_experiment.tex
\begin{figure*}[htp]
  \begin{center}
\includegraphics[width=0.6\textwidth]{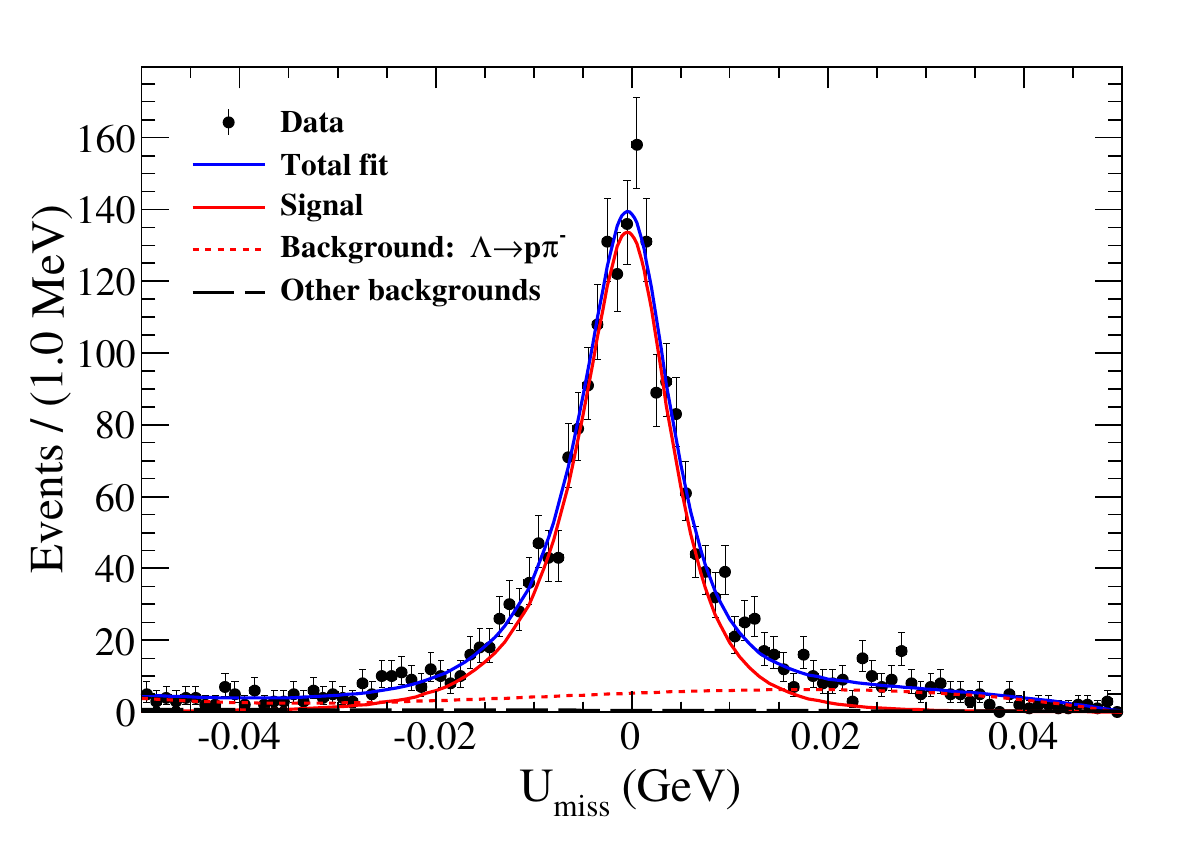}
    \caption{\textbf{The $\bm{U_{\rm miss}}$ distribution of $\bm{\Lambda \to p e^{-}\bar{\nu}_{e}}$ and $\bm{\bar{\Lambda}\to\bar{p} e^{+}\nu_{e}}$ candidates}. The black points with error bars represent the observed number of events in each bin, with statistical uncertainties. The blue solid line shows the total fit, which includes the signal candidates (red solid), the dominant background (red dotted line), and the other background contributions (black long-dashed line).}
    \label{fig:fit_bf}
  \end{center}
\end{figure*}

 Our results are based on the $(10.087\pm0.044)\times10^9$ $J/\psi$ events collected with the multi-purpose detector BESIII~\cite{BESIII:2021cxx}. 
The $J/\psi$ resonance decays into a $\Lambda \bar{\Lambda}$ pair with a probability of $(1.89\pm0.09)\times 10^{-3}$~\cite{ParticleDataGroup:2024cfk}. All final-state particles except the neutrino are electrically charged and can therefore be reconstructed in the multilayer drift chamber, where a superconducting solenoid provides a magnetic field enabling momentum determination with an accuracy of $0.5\%$ at 1.0~GeV/c. The semi-leptonic and hadronic $\Lambda$($\bar{\Lambda}$) candidates are identified by combining $pe^- (\bar{p}e^+)$ pairs and $\bar{p}\pi^+ (p\pi^-)$, respectively. To obtain electrons/positrons with sufficient quality, their momenta were required to be larger than 0.1~GeV/c$^2$. 
The semi-leptonic decays are identified using the variable $U_{\rm miss} = E_{\rm miss} - c\cdot |\vec{p}_{\rm miss}|$, where $E_{\rm miss}$ and $\vec{p}_{\rm miss}$ denote the missing energy and momentum of the semi-leptonic decay, respectively. For candidates of semi-leptonic decay, the $U_{\rm miss}$ distribution is uniformly distributed around zero, while the hadronic $\Lambda$ background features a right-skewed distribution which peaks at $U_{\rm miss}\sim 0.022$~GeV. The signal yield is determined from an extended unbinned maximum likelihood fit~\cite{Verkerke:2003ir} of the $U_{\rm miss}$ distribution, shown in Figure~\ref{fig:fit_bf}, which combines both $\Lambda\to pe^-\bar{\nu}_e$ and $\bar{\Lambda}\to \bar{p}e^+\nu_e$. 
The signal shape is described using Monte Carlo (MC) simulation convoluted with a Gaussian function to account for the resolution differences between data and MC. The shape of the dominant background contribution, $\Lambda\to p \pi^-$, is obtained using the corresponding MC sample, while the rest of the background contribution is described by a linear function. The yields and the parameters of the Gaussian and linear functions are floated in the fit.
More details of the selection and analysis procedure are given in Methods.

For the determination of the semi-leptonic absolute branching fraction, we employ the double-tagging (DT) technique, as pioneered in the MARK-III experiment~\cite{MARK-III:1985hbd}. A double tag event means that one $\Lambda$ is reconstructed as $\Lambda\to p e \bar{\nu}_e$ and the antihyperon partner as $\bar{\Lambda}\to \bar{p}\pi^+$. The technique also requires knowledge of the single tag sample, which means that only one side is reconstructed as $\bar{\Lambda}\to\bar{p}\pi^+$. It has the additional advantage that the four momentum transfer $q^2$ can be reconstructed unambiguously~\cite{CLEO:2008bkh}. A more detailed account of the technique is provided in Methods. The branching fraction of the semi-leptonic decay of the $\Lambda$ is determined from 

\begin{equation}
\label{eq_bf}
{\mathcal B}(\Lambda\to pe^-\bar{\nu}_e)= \frac{N_{\rm DT}}{\epsilon_{\rm DT}}\frac{\epsilon_{\rm ST}}{N_{\rm ST}},
\end{equation}
where $N_{\rm DT}$ and $\epsilon_{\rm DT}$ are the semi-leptonic event yield and detection efficiency in the DT sample, respectively, while $N_{\rm ST}$ and $\epsilon_{\rm ST}$ are the event yield and detection efficiency obtained using the ST sample. 
The ST yield and efficiency are determined using the same methods from Ref.~\cite{BESIII:2021ynj}, where they are obtained from a fit to the invariant mass distribution of the ST candidate distribution. These are found to be $N_{\rm ST}=(14.328\pm0.005)\times10^6$ and $\epsilon_{\rm ST}=(54.09 \pm 0.01)\%$, respectively.
The signal event yield, $N_{\rm DT}=1854\pm49$, is obtained from the fit in Fig.~\ref{fig:fit_bf}. The DT detection efficiency, $\epsilon_{\rm DT}=(8.58\pm0.01)\%$, is determined from a dedicated simulated sample including the production and decay dynamics of the full process. Equation~\eqref{eq_bf} yields $\mathcal{B}(\Lambda\to p e^- \bar{\nu}_{e})=(8.16\pm0.22\pm0.15)\times 10^{-4}$ where the first uncertainty is statistical and the second is systematic, discussed in Methods. This is in agreement with the world average $\mathcal{B}(\Lambda\to p e^- \bar{\nu}_{e})=(8.34\pm0.14)\times 10^{-4}$~\cite{ParticleDataGroup:2024cfk}.

Whereas the branching fraction determination requires only yields and efficiencies, the extraction of the form factors requires a detailed understanding of the decay kinematics. To increase the sample purity, we retain only events with $|U_{\rm miss}|<0.02$~GeV, resulting in $1791\pm47$ signal and $207\pm30$ background events. From these events, we reconstruct the kinematic variables $\xi$ and determine the transition form factors in $\omega$ using an unbinned maximum log-likelihood fit that accounts for the multidimensional reconstruction efficiency. In contrast to previous measurements~\cite{Wise:1980iq, Bristol-Geneva-Heidelberg-Orsay-Rutherford-Strasbourg:1983jzt, Dworkin:1990dd}, we can exploit the fact that the $\Lambda$ is produced along with a $\bar{\Lambda}$ to our advantage. On an event-by-event basis, it allows us to fully reconstruct the kinematics of the semi-leptonic decay, while globally, it provides detailed information on the production mechanism of the quantum-entangled $\Lambda-\bar{\Lambda}$ pair. The $\Lambda$ polarization and the spin correlations between the pair are uniquely determined by the two production parameters, $\alpha_{\psi}$ and $\Delta\Phi$. These, along with the two decay parameters, $\alpha_-$ and $\alpha_+$, have already been precisely measured using the same data set in Refs.~\cite{BESIII:2018cnd, BESIII:2022qax}. Therefore, the free physics parameters in $\omega$ are reduced to three from six -- the couplings $g_{av}$, $g_w$ and $g_{av2}$. These couplings can therefore be determined with optimal precision, even with a relatively modest data sample. 
The details of the maximum log-likelihood fit procedure and the systematic uncertainties are described in Methods.

%% file: 05_results.tex
The final results for the axial-vector and weak-magnetism couplings are summarized in Table~\ref{tab:result}. They are determined separately for the $e^{\pm}$ semi-leptonic channels and simultaneously
by assuming $g_i^- = -g_i^+$, $i=\{av,av2\}$, and $g_w^- = g_w^+$. 
Two results are provided: one where the weak-electricity coupling is fixed to the SU(3) symmetry limit~\cite{Cabibbo:2003cu, Cabibbo:2003ea}, $g_{av2}=0$, and the other where it is determined along with the other two couplings. The first result allows for a more precise extraction of $g_{av}$ and $g_{w}$. The value of the weak-magnetism coupling, $\langle g_w\rangle =0.89\pm{0.38}$, is consistent with both the conserved vector current hypothesis prediction of $0.97$~\cite{Gell-Mann:1960mvl} and previous measurement of $g_w=0.15\pm0.30$~\cite{Dworkin:1990dd}, with the latter agreement at the $1.5\sigma$ level.
The uncertainty achieved for $\langle g_w\rangle$ in our analysis is comparable to that of the Fermilab measurement~\cite{Dworkin:1990dd}. 
For the second set, the weak-electricity coupling is found to be $\langle g_{av2}\rangle=-0.19^{+0.65}_{-0.63}{\pm0.18}$, providing the first measurement for $\Lambda$ semi-leptonic decay. It is also three times more precise compared to the KTeV determination of g$_{av2}$ for the $\Xi^0\to\Sigma^+$ semi-leptonic decay~\cite{KTeV:2001djr}. The corresponding average axial-vector and weak-magnetism couplings are $\langle g_{av}\rangle=0.706^{+0.089}_{-0.086}$ and $\langle g_w\rangle=0.77^{+0.53}_{-0.49}$, consistent with the first set of results within the uncertainties; both are also compatible with the corresponding lattice QCD calculation~\cite{Bacchio:2025auj}.

\begin{table*}[htp]
  \begin{center}
    \caption{\textbf{Summary of the results} }
    \renewcommand\arraystretch{1.2}
    \setlength{\tabcolsep}{3mm}{
      \begin{tabular}{llll}
        \hline\hline
         Observable&									This work&								Previous result & Refs.\\
        \hline
        ${\mathcal B}(\Lambda\to pe^-\bar{\nu}_{e})$&		$\phantom{-}(8.16\pm0.22\pm 0.15)\times10^{-4}$&			$(8.34\pm0.14)\times10^{-4}$ & \cite{ParticleDataGroup:2024cfk}\\
        $g_{av}^-$&$\phantom{-}0.742^{+0.075}_{-0.057}\pm0.009$&$0.718\pm0.015$ & \cite{ParticleDataGroup:2024cfk}\\
        $g_{av}^+$&									$-0.706^{+0.073}_{-0.069}\pm0.014$&$-$&\\
        $\langle g_{av}\rangle$ & $\phantom{-}0.729^{+0.048}_{-0.047}\pm0.007$&$-$&\\
        $g_{w}^-$&$\phantom{-}0.93\pm{0.51}\pm0.17$&$0.15\pm0.30$&\cite{Dworkin:1990dd}\\
        $g_{w}^+$&$\phantom{-}0.89\pm{0.49}\pm0.20$&$-$&\\
        $\langle g_w\rangle$&$\phantom{-}0.89\pm{0.35}\pm0.14$&$-$&\\
        \hline
        $|V_{us}|_\textrm{SU(3)}$&$\phantom{-}{0.2194}\pm0.0094$&\multirow{2}{*}{$0.2224\pm0.0034$}&\multirow{2}{*}{\cite{Cabibbo:2003cu,Cabibbo:2003ea}}\\
        $|V_{us}|_{\rm LQCD}$&{$\phantom{-}0.2339\pm0.0041$}& &\\
        $|V_{us}|\cdot\sqrt{f_1^2+3g_1^2}$&{$\phantom{-}0.4496\pm0.0075$}&$-$&\\
        \hline
        $\langle g_{av}\rangle$&$\phantom{-}0.706^{+0.089}_{-0.086}{\pm0.026}$&$-$ & \\
        $\langle g_w\rangle$&$\phantom{-}0.77^{+0.53}_{-0.49}{\pm0.14}$&$-$&\\
        $\langle g_{av2}\rangle$&$-0.19^{+0.65}_{-0.63}{\pm0.18}$&$-$ & \\
        \hline\hline
        \multicolumn{4}{p{12cm}}{\footnotesize The branching fraction $\mathcal{B}$ of  $\Lambda\to pe^-\bar{\nu}_e$ and form factor ratios for $\Lambda\to pe^-\bar{\nu}_e (g_{av}^-, g_w^-)$, $\bar{\Lambda}\to\bar{p}e^+\nu_e (g_{av}^+, g_w^+)$ and the average $\langle g_{av}\rangle$ and $\langle g_w\rangle$ with the constraint that $g_{av2}=0$, where the first and second uncertainties are statistical and systematic, respectively;
        the value of $|V_{us}|$ within the SU(3) symmetry limit and with the lattice QCD prediction, the product of the $|V_{us}|$ and the form factors $|V_{us}|\cdot\sqrt{f_1^2+3g_1^2}$, where the uncertainties are total uncertainties calculated as a sum in quadrature of the individual uncertainties; the average $\langle g_{av}\rangle$ and $\langle g_w\rangle$ with the average $\langle g_{av2}\rangle$ result with statistical and systematic uncertainties, respectively. 
        }\\
      \end{tabular}
    }\label{tab:result}
  \end{center}
\end{table*}

To extract the CKM matrix element $|V_{us}|$ from equation~(\ref{eq:Vusf1}) we use the $\Lambda$ lifetime $\tau_{\Lambda}=(2.617\pm0.010)\cdot10^{-10}$~s~\cite{ParticleDataGroup:2024cfk}, and our determined $\langle g_{av}\rangle$, $\langle g_w\rangle$ and $\mathcal{B}(\Lambda\to pe^-\bar{\nu}_e)$. 
Meanwhile, we assume $g_2=0$~\cite{Cabibbo:2003cu},
which is consistent with our second set of form factor results and also assumed by the previous experimental measurements~\cite{Dworkin:1990dd, Bristol-Geneva-Heidelberg-Orsay-Rutherford-Strasbourg:1983jzt, Wise:1980xx} 
and for the determination of $|V_{us}|$~\cite{Cabibbo:2003cu,Cabibbo:2003ea}. The term ${\mathcal O}(\beta^3)$ is ignored as its contribution is negligible within current experimental accuracy.
For the decay rate, one also needs to account for radiative effects, which modify the decay rate by $(1.6\pm0.5)\cdot10^{-2}$~\cite{Flores-Mendieta:1996vgq}.

The final hurdle before obtaining $|V_{us}|$ is that either $f_1$ or $g_1$ must be explicitly provided, either from theory or lattice QCD. This is similar to kaon semi-leptonic decays, where the corresponding form factor $f_{+}$ is required as input~\cite{ParticleDataGroup:2024cfk}. 
As a reference value for our discussions, we assume that flavour SU(3) symmetry is conserved. Under this assumption, 
$f_1$ takes on the value $-\sqrt{3/2}$~\cite{Cabibbo:2003cu} and $|V_{us}|_\textrm{SU(3)}={0.2194}\pm0.0036_{\rm BESIII~BF}\pm0.0087_{\rm BESIII~FF}\pm0.0004_{\tau_{\Lambda}}\pm0.0005_{\rm RC}$. The first and second uncertainties are from our measurements of the branching fraction and form factors, respectively, the third from the $\Lambda$ lifetime, while the fourth is the radiative correction uncertainties~\cite{Flores-Mendieta:1996vgq}.
This result is in good agreement with the value extracted by Cabibbo from the $\Lambda\to pe^-\bar{\nu}_e$ world average, $|V_{us}|=0.2224\pm0.0034$~\cite{Cabibbo:2003cu}. 
The value is also consistent with the CKM unitarity requirement within $0.9$ standard deviations, $|V_{us}|=0.22799\pm0.00137$~\cite{ParticleDataGroup:2024cfk, HFLAV:2022esi}.
Various theoretical models account for the potential effects of flavor SU(3) symmetry-breaking, therefore we require a more accurate input that reflects this fact. The recent lattice QCD result~\cite{Bacchio:2025auj} provides the first detailed determination of the relevant form factors. Using these values together with our measured branching fraction, we obtain $|V_{us}|_{\rm LQCD}=0.2339\pm0.0038_{\rm BESIII~BF}\pm0.0004_{\tau_{\Lambda}}\pm0.0006_{\rm RC}\pm0.0011_{\rm LQCD}$, where the fourth uncertainty arises from the lattice QCD calculation.  

\begin{figure}[htp]
  \begin{center}
      \includegraphics[width=1.0\textwidth]
      {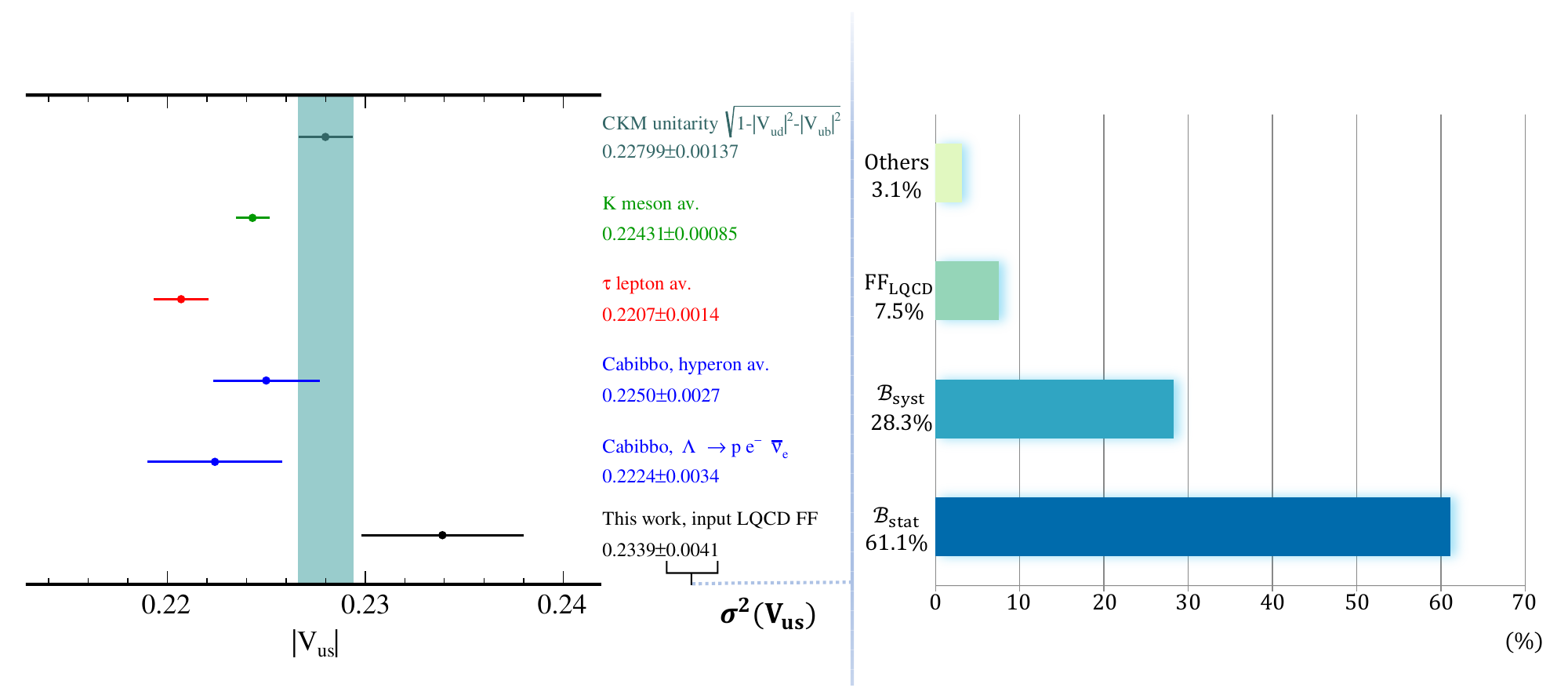}
      \caption{\textbf{The $\bm{|V_{us}|}$ result.} (left) The first-row CKM unitarity constraint and $K$ average are from Ref.~\cite{ParticleDataGroup:2024cfk}. The $\tau$ average is from the Heavy Flavor Averaging Group~\cite{HFLAV:2022esi}. The two hyperon results were evaluated by Cabibbo in~\cite{Cabibbo:2003ea, Cabibbo:2003cu}. (right) The contributing sources to the $|V_{us}|$ uncertainty, in terms of the variance.}  
      \label{fig:Vus}
    \end{center}
\end{figure}

The individual contributions to the uncertainty of this measurement are shown in Fig.~\ref{fig:Vus}. By using lattice-determined form factors, the corresponding contribution to the $|V_{us}|$ uncertainty is reduced from 84.9\% to 7.5\%. While further improvements in the form factor remain, the statistical uncertainty of the branching fraction measurement, 0.0032, is now the dominant source. This demonstrates how the interplay between experimental measurements and lattice QCD calculations can enable increasingly precise determinations of CKM matrix elements. The combined BESIII and lattice QCD result, $|V_{us}|_{\rm LQCD}$, is consistent with CKM unitarity at the $1.4\sigma$ level, but is approximately two standard deviations larger than the value predicted by Cabibbo~\cite{Cabibbo:2003cu, Cabibbo:2003ea}, as shown in Fig.~\ref{fig:Vus}. 

For future comparisons, we also provide an experimentally based result that does not rely on any theoretical form factor input.
The product of $|V_{us}|$ with the dominant form factor contribution at zero momentum transfer, $q^2=0$, is extracted as $|V_{us}|\cdot\sqrt{f_1^2+3g_1^2}=0.4496\pm0.0061_{\rm BESIII~BF_{stat}}\pm0.0041_{\rm BESIII~BF_{syst}}\pm0.0009_{\tau_{\Lambda}}\pm0.0011_{\rm RC}$. The reason is that after assuming $g_2=0$ and neglecting the $\mathcal{F}_2$ contribution, as the kinematic pre-factor is small, $\frac27\beta^2\approx0.007$, the partial decay width $\Gamma$ is approximately proportional to 
$V_{us}^2({f_1^2+3g_1^2})$. 
Under these two assumptions, the short-handed formula becomes identical to the standard next-to-leading order form in the LQCD framework~\cite{Bacchio:2025auj}. As a technical side note, the radiative corrections applied to the weak decay constant differ between $|V_{us}|_{\text{SU(3)}}$ and $|V_{us}|_{\text{LQCD}}$. Consequently, when applying radiative corrections to the $|V_{us}|\cdot\sqrt{f_1^2+3g_1^2}$, we adopt the same correction method as used for $|V_{us}|_{\text{LQCD}}$. 
This product is model-independent, as it reduces reliance on theory-based form factor inputs and thus provides a stringent constraint for validating theoretical descriptions of semi-leptonic baryon decays. To our knowledge, this is the first determination of such a product in any semi-leptonic baryon decay. \\

%% file: 06_conclusion.tex
Summarizing our findings, using spin-entangled and polarized strange baryon--antibaryon pairs, we have reported the first absolute branching fraction determination for the semi-leptonic $e$-mode of strange baryon decays. We have determined the parameters related to the internal structure of the $\Lambda$ baryon --- the first measurement of the weak-electricity coupling $g_{av2}$, as well as the axial-vector coupling $g_{av}$ and the weak-magnetism coupling $g_{w}$ in $\Lambda$ semi-leptonic decay. We obtain two values for $|V_{us}|$: one uses the SU(3) symmetry-preserved vector form factor value as normalization~\cite{Cabibbo:1963yz}, and the other uses form factor input from lattice QCD calculations~\cite{Bacchio:2025auj}. This is the first experimental determination for the $|V_{us}|$ after a more than thirty-year break for the study of semi-leptonic $\Lambda$ decay. 
The novel modular approach~\cite{Batozskaya:2023rek} is applied in the analysis, marking the first exploitation of polarization and quantum entanglement in baryon semi-leptonic decays. This innovative methodology significantly enhances the single‑event sensitivity, as demonstrated by a direct comparison with the Fermilab measurement~\cite{Dworkin:1990dd}: although their dataset is approximately twenty times larger than ours, we achieve a comparable precision for $g_{w}$, while the uncertainty on $g_{av}$ is only about a factor of 2.5 larger, and, most notably, we determine $g_{av2}$ for the first time, a parameter previously inaccessible. 
Along with the first exploitation of polarization and quantum entanglement in baryon semileptonic decays, this work establishes the foundation for a systematic research program. The same method is already being extended to our following studies of other baryon semileptonic decays at BESIII~\cite{BESIII:2025wov} and applies to both present and future particle‑antiparticle collider facilities such as the upcoming PANDA experiment~\cite{PANDA:2020zwv} and the proposed Super Tau‑Charm Factories~\cite{Achasov:2023gey, Barnyakov:2020vob}. Because form factors in different semileptonic decays are interrelated~\cite{Cabibbo:2003cu}, results obtained with this method can be combined across channels. Together with continued advances in lattice QCD~\cite{Bacchio:2025auj}, this program offers a clear path toward a model‑independent determination of $|V_{us}|$ with precision competitive with that from kaon decays, thereby providing a critical and independent test of CKM unitarity.

%% file: 07_methods.tex
\section*{Methods\label{sec:meth}}
\noindent
\textbf{Experimental Apparatus}\\ The BESIII detector~\cite{BESIII:2009fln} records symmetric $e^+e^-$ collisions provided by the BEPCII storage ring~\cite{Yu:2016cof}. In this cylindrical system, tracks of charged particles in the detector are reconstructed from track-induced signals, and their momenta are determined from the track curvature in the multilayer drift chamber~(MDC). The flight time of the charged particles is recorded by a plastic scintillator time-of-flight subdetector (TOF). Electromagnetic showers from photons are reconstructed in the electromagnetic calorimeter~(EMC). More details about the design and performance of the BESIII detector are found in Ref.~\cite{BESIII:2009fln}. 

{\setlength{\parskip}{15pt}
\noindent
\textbf{Monte Carlo Simulation}\\ For the selection, efficiency determination, and background evaluation of the single and double tag processes, Monte Carlo (MC) simulations have been used. The {\sc geant4}-based package~\cite{GEANT4:2002zbu} includes the geometric description and detector response of the BESIII detector. The inclusive MC sample includes both the production of the $J/\psi$
resonance and the continuum processes incorporated in the {\sc kkmc} simulation package~\cite{Jadach:2000ir}. All particle decays are modelled with {\sc evtgen}~\cite{Lange:2001uf} using the branching fractions from the Particle Data Group when available~\cite{ParticleDataGroup:2024cfk}, or else estimated with {\sc lundcharm}~\cite{Yang:2014vra}. Final state radiation of the charged final state particles is simulated with {\sc photos}~\cite{Richter-Was:1992hxq}. The inclusive MC sample does not consider the decay dynamics and is thus only used for qualitative background studies. For the signal channel $J/\psi\to\Lambda(\to pe^-\bar{\nu}_e) \bar{\Lambda}(\to\bar{p}\pi^+) + c.c.$ and the background, $J/\psi\to\Lambda(\to p\pi^-)\bar{\Lambda}(\to\bar{p}\pi^+)$ dedicated MC samples were produced. In addition, these two modes include realistic production and decay dynamics using the physics parameter values in agreement with the results in Table~\ref{tab:result}, and in Ref.~\cite{BESIII:2018cnd}. 

{\setlength{\parskip}{15pt}
\noindent
\textbf{Event selection criteria} \\
For the single-tagged $\bar{\Lambda}(\Lambda)$ sample, the same selection procedure is applied as in Ref.~\cite{BESIII:2021ynj}.
The double tag candidates are selected from the remaining tracks recoiling with respect to the ST candidates. All charged tracks must satisfy $|\cos \theta_{\rm LAB}| < 0.93$, where $\theta$ is defined with respect to the $z$-axis,
which is the symmetry axis of the MDC. Only events with four charged tracks are considered further. For the $\Lambda$ reconstruction, the pair of charged tracks is constrained to originate from a common vertex. Furthermore, the decay length of the $\Lambda$ candidate is required to be greater than zero. To identify whether a charged track is an electron/positron or a pion, particle identification is required. Each track is assigned a likelihood based on the MDC ionization energy loss and information from the TOF and EMC sub-systems.
The charged track on the DT side, other than the electron/positron, is assumed to be a proton. 
After these requirements have been imposed on the data set, there is still a sizable contribution from the non-leptonic $\Lambda\to p\pi^-$ decay. To suppress these events, we assume that the DT semi-leptonic candidates are in fact the hadronic decay but with a misidentified electron. This modifies the $\Lambda$ decay vertex, which in turn also modifies the $\Lambda$ momentum vector.
Then a four constraint kinematic fit (4C-fit) is imposed, which tests energy and momentum conservation in the $J/\psi\to\Lambda\bar{\Lambda}$ hypothesis. If the $\chi^2$ of the 4C-fit is lower than fifty, then the event is more likely to be a background and hence discarded. The $\chi^2_{4C}$ selection removes $(84.4\pm0.2)\%$ and $(4.30\pm0.02)\%$ of the background and signal components, respectively.  
Except for $\Lambda\to p\pi^-$, we find no other source of peaking background. A potential background that includes an extra photon, $\Lambda\to p\pi^- \gamma$ decay, was also studied with an exclusive MC simulation. Its contribution was found to be negligible, only 0.02\%. 
\\ To improve the data quality, the electron momentum is required to be larger than 0.1 GeV/$c$. Since the neutrino is undetected, the kinematic variable $U_{\rm miss} \equiv E_{\rm miss}-c|\vec{p}_{\rm miss}|$, is introduced. Here, $E_{\rm miss}$ and $\vec{p}_{\rm miss}$ are the missing energy and momentum carried by the neutrino, respectively. These are calculated from $E_{\rm miss}=E_{\rm beam}-E_{p}-E_{e^-}$ and $\vec{p}_{\rm miss}=\vec{p}_{\Lambda}-{\vec p}_{p}-{\vec p}_{e^-}$, respectively, where $E_{\rm beam}$ is the beam energy, $E_{p(e^-)}$ and ${\vec p}_{p(e^-)}$ are the measured energies and momentum of the proton (electron), respectively. To determine the $\Lambda$ momentum, $\vec{p}_{\Lambda}$, the momentum and magnitude from the produced $\bar{\Lambda}$ partner combined with the known beam energy are used.
For the signal candidates, the $U_{\rm miss}$ distribution is expected to peak around zero. The four-momentum transfer $q^2$ is determined as $q^2 = (E_{\rm beam}-E_{p})^2-(\vec{p}_{\Lambda}-{\vec p}_{p})^2$.

{\setlength{\parskip}{15pt}
\noindent
\textbf{Double-tagging technique} \\
At BESIII, semi-leptonic events are produced in the reaction $e^+e^-\to J/\psi \to \Lambda\bar{\Lambda}$. For each event, we first reconstruct the decay $\bar{\Lambda}\to \bar{p} \pi^+$ (or $\Lambda\to p \pi^-$), which defines the single-tag (ST) sample. The semi-leptonic decay is then searched for in the system recoiling against the ST candidate; events in which it is identified from the double-tag (DT) sample.
The ST and DT event yields are given by
\begin{equation}
N_{\mathrm{ST}} = 2 N_{\Lambda\bar{\Lambda}} \mathcal{B}_{\mathrm{ST}} \epsilon_{\mathrm{ST}}, 
\label{eq:st}
\end{equation}
and 
\begin{equation}
N_{\mathrm{DT}} = 2 N_{\Lambda\bar{\Lambda}} \mathcal{B}_{\mathrm{ST}} \mathcal{B}(\Lambda\to pe^-\bar{\nu}_e) \epsilon_{\mathrm{DT}},
\label{eq:dt}
\end{equation}
where $N_{\Lambda\bar{\Lambda}}$ is the number of produced $\Lambda\bar{\Lambda}$ pairs, $\mathcal{B}_{\mathrm{ST}}$ and $\mathcal{B}(\Lambda \to p e^- \bar{\nu}e)$ are the corresponding branching fractions, and $\epsilon_{\mathrm{ST}(\mathrm{DT})}$ is the ST(DT) reconstruction efficiency. The common factor $2 N_{\Lambda\bar{\Lambda}} \mathcal{B}_{\mathrm{ST}}$ cancels in the ratio, allowing the absolute branching fraction to be determined directly from the measured yields and efficiencies,
\begin{equation} 
{\mathcal B}(\Lambda\to pe^-\bar{\nu}_e)= \frac{N_{\rm DT}}{\epsilon_{\rm DT}}\frac{\epsilon_{\rm ST}}{N_{\rm ST}}.
\end{equation}


{\setlength{\parskip}{15pt}}
\noindent
\textbf{Definition of the helicity amplitudes}\\  
In the $e^+e^-\to\Lambda(\to pe^-\bar{\nu}_e)\bar{\Lambda}(\to\bar{p}\pi^+)$ process, the
`master coordinate system', denoted $\mathcal{R}$, is defined in the $e^+e^-$ centre-of-momentum system. In this system, we define the unit vector $\hat{\vec z}$ in the direction of the positron momentum. The coordinate system $\mathcal{R}_{\Lambda}$ is then defined in the rest frame of the $\Lambda$ baryon, with the $z$ axis along the unit vector $\hat{\vec z}_{\Lambda}$ defined by the direction of the $\Lambda$ momentum in the $\mathcal{R}$ system. A Cartesian coordinate system with $\hat{\vec x}_{\Lambda}$ and $\hat{\vec y}_{\Lambda}$ unit vectors is defined as
\begin{equation}\label{eq:Cartesiancoor}
\hat{\vec x}_{\Lambda}=\frac{\hat{\vec z}\times\hat{\vec z}_{\Lambda}}{|\hat{\vec z}\times\hat{\vec z}_{\Lambda}|}\times\hat{\vec z}_{\Lambda},\quad\hat{\vec y}_{\Lambda}=\frac{\hat{\vec z}\times\hat{\vec z}_{\Lambda}}{|\hat{\vec z}\times\hat{\vec z}_{\Lambda}|}.
\end{equation}

The helicity system $\mathcal{R}_{\bar{\Lambda}}$ is defined in the same way in the $\bar{\Lambda}$ rest frame, and because $\hat{\vec z}_{\Lambda}=-\hat{\vec z}_{\bar{\Lambda}}$.

{\setlength{\parskip}{15pt}
\noindent
\textbf{The maximum log-likelihood fit procedure}\\ 
A simultaneous fit is performed to the two c.c. channels, $J/\psi \to \Lambda(\to pe^-\bar{\nu}_e)\bar{\Lambda}(\to \bar{p}\pi^+)$ and $J/\psi \to \Lambda(\to p\pi^-)\bar{\Lambda}(\to \bar{p}e^+\nu_e)$. At this level of precision charge parity~(CP) is a conserved symmetry, so these CP-odd relations are valid: $\alpha_-=-\alpha_+$, $g_{av}^-=-g_{av}^+$, $g_{w}^-=g_{w}^+$ and $g_{av2}^-=-g_{av2}^+$ which reduces the number of free parameters in $\bm{\omega}$. For $N$ number of experimental events, the likelihood, constructed from the probability density function for an event characterized by $\bm{\xi}_{i}$ is}
\begin{eqnarray} \begin{aligned}
  \mathcal{L}=\prod_{i=1}^{N} \mathcal{P}\left(\bm{\xi}_{i}; \bm{\omega}\right)=\prod_{i=1}^{N} \frac{\mathcal{W}\left(\bm{\xi}_{i}; \bm{\omega}\right) \epsilon\left(\bm{\xi}_{i}\right)}{\mathcal{N}\left(\bm{\omega}\right)},
  \label{likelihood function}
\end{aligned}\end{eqnarray}
where $\epsilon\left(\bm{\xi}_{i}\right)$ is the detection efficiency, the normalization factor $\mathcal{N}\left(\bm{\omega}\right)=\int \mathcal{W}\left(\bm{\xi}; \bm{\omega}\right) \epsilon\left(\bm{\xi}\right) d\bm{\xi}$ with weights $\mathcal{W}\left(\bm{\xi}; \bm{\omega}\right)$ as specified in Eq.~\eqref{eq:W}.
The normalization factor is approximated as $\mathcal{N}(\bm{\omega})=\frac{1}{M}\sum_{j=1}^{M}[\mathcal{W}(\bm{\xi}_j;\bm{\omega})/\mathcal{W}(\bm{\xi}_j;\bm{\omega_{\rm gen}})]$ using $M$ Monte Carlo events $\bm{\xi}_j$ generated using the angular distribution of the full process~\eqref{eq:W} with fixed parameters $\bm{\omega_{\rm gen}}$, propagated through the detector and reconstructed in the same way as the experimental data. $M$ is chosen to be much larger than $N$; in this case the ratio is $M/N=470$.
To take into account the difference in detection efficiency between MC and data, the detection efficiency is corrected for the final state particles, $p$, $\bar{p}$, $\pi^+$, $\pi^-$, $e^+$, and $e^-$. Applying the same method as described in Ref.~\cite{ele_sys}, the correction factors are obtained using the control samples $J/\psi\to p\bar{p}\pi^+\pi^-$ and $e^+e^-\to e^+e^-\gamma$, where the correlation between the charged particle momentum and its polar angle acceptances is taken into account. Form factors can be expressed in different parameterizations. Our final results are quoted in the standard Weinberg classification, although an equivalent helicity-based representation also exists~\cite{Feldmann:2011xf}. In the Weinberg basis, the axial-vector and weak-electricity couplings exhibit strong correlation, {$\rho(g_{av},g_{av2})\geq95\%$}. To reduce this in the fit, we instead use the helicity form factors $g_+$, $g_{\perp}$, $f_+$, and $f_{\perp}$ in the joint angular distribution, 
and subsequently transform them to the Weinberg representation. At zero momentum transfer, this relation is
\begin{equation}\label{eq:helFF_WFF}
    \begin{aligned}
    f_+ = &\,\, f_1, &  &f_{\perp} =  f_1+(2-\beta)f_2,\\
    g_+ = & \,\,g_1, &  & g_{\perp} =  g_1-\beta g_2.\\
    \end{aligned}
\end{equation}
This procedure leads to slightly asymmetrical statistical uncertainties for the extracted form factor ratios (Table~\ref{tab:result}). 
For $g_w^-$, $g_w^+$, and $\langle g_w\rangle$ in the first set, this asymmetry is not visible at the quoted precision. For the final $|V_{us}|$ determination, the effect of these asymmetric uncertainties is negligible; we therefore quote a symmetric uncertainty.
\\
To determine the parameters, the Minuit package from the CERN library is used~\cite{James:1975dr}. 
The minimizing function is given by $S=-\ln \mathcal{L}_{\text {data}} + \ln \mathcal{L}_{\text {bkg}}$, where $\mathcal{L}_{\text {data}}$ and $\mathcal{L}_{\text {bkg}}$ represent the two likelihoods for the semi-leptonic modes and the background events. The background contribution is evaluated using simulated data.
In addition, the operational conditions were slightly different for the four data-taking periods (2009, 2012, 2017–2018, and 2018–2019),  most notably in the nominal value of the magnetic field. For this reason, the likelihoods are evaluated separately for the four different run periods.

{\setlength{\parskip}{15pt}
\noindent
\textbf{The $\bm{|V_{us}}|$ determination}\\ 
The partial decay width for $\Lambda\to pe^-\overline{\nu}_e$ decay, given in Eq.~\eqref{eq:Vusf1}, 
has been used in the determination of $|V_{us}|$ using Cabibbo method~\cite{Cabibbo:2003ea,Cabibbo:2003cu,Garcia:1985xz}, where the $q^2$ dependence of the form factors is not included. To test the systematic effects associated with this approximation, we include the ${\cal O}(q^2)$ terms in the expansion of the relevant form factors about $q^2=0$, following Eq.~(A11) of Ref.~\cite{Bacchio:2025auj}:
\begin{equation}\label{eq:q2_depen}
    \begin{aligned}
    f_i(q^2) \approx & f_i\cdot\left(1+\frac{\braket{r_{f_i}^2}}{6}q^2\right)\quad\text{ and }\quad
    g_j(q^2) \approx  g_j\cdot\left(1+\frac{\braket{r_{g_j}^2}}{6}q^2\right)\ ,
    \end{aligned}
\end{equation}
where the parameters $\braket{r_{f_i}^2}$ and $\braket{r_{g_j}^2}$ are defined consistently for both the dipole parametrization~\cite{Kadeer:2005aq,Korner:1976hv} and the so called \textit{z-expansion} used in the lattice QCD calculation~\cite{Bacchio:2025auj} and in the quasipotential approach~\cite{Faustov:2018dkn}, although their numerical values differ between these approaches. For the dipole parameterization, these parameters are given by:
 \begin{equation}\label{eq:rad}
 \begin{aligned}
    \braket{r_{f_i}^2}= & 6{\sum_{n=0}^{i}\frac{1}{m^2_V+n\alpha_R^{-1}}}\quad\text{ and }\quad
    \braket{r_{g_j}^2}= & 6{\sum_{n=0}^{j}\frac{1}{m^2_A+n\alpha_R^{-1}}},\
    \end{aligned}
 \end{equation}
with dominant pole masses $m_V=m_{K^{\ast}(892)^0}=0.892$~GeV and $m_A=m_{K_1(1270)}=1.273$~GeV~\cite{Korner:1976hv}. Here, $K^{\ast}(892)^0$ and $K_1(1270)$ are the lowest-lying strange vector mesons with quantum numbers $J^P=1^-$ and $1^+$, respectively. 
The slope of the Regge trajectory is taken as $\alpha_R=0.9$~GeV$^{-2}$.

The decay width, which takes into account radiative corrections, is given by ~\cite{Flores-Mendieta:1996vgq}
\begin{equation}\label{eq:rc}
\Gamma^c= \Gamma \cdot (1+Cr).
\end{equation}
Here $Cr$ is the radiative correction factor to the decay rate, and it has been calculated to be $Cr=(1.6\pm0.5)\%$ for the decay $\Lambda\to p e^-\bar{\nu}_e$~\cite{Flores-Mendieta:1996vgq}. 

We determine $|V_{us}|$ in two ways. In the first approach, Eq.~(\ref{eq:Vusf1}) is evaluated using our measured form factors $\langle g_{av}\rangle$ and $\langle g_{w}\rangle$, while fixing $f_{1}$ and $g_{av2}$ to their SU(3)-conserving values, $-\sqrt{3}/2$ and $0$, respectively. Although the universal, nuclear-structure-independent electroweak radiative correction has been extensively studied in Refs.~\cite{Marciano:1985pd,Marciano:2005ec,Czarnecki:2019mwq,Seng:2018yzq,Shiells:2020fqp,Seng:2020wjq,Ma:2023kfr,Hayen:2020cxh}, such studies for hyperon decays are very limited. The electroweak radiative correction is incorporated through the prescription $G_F^2\equiv G_F^2\cdot(1+\Delta_R^V) = G_F^2\cdot(1+0.0191)$, following Refs.~\cite{Marciano:1985pd,Marciano:2005ec,Czarnecki:2019mwq}, replacing the nucleon contribution to $\Delta_R^V$ with that of the $\Lambda$ hyperon. In the second approach, $|V_{us}|$ is determined using lattice-QCD form factors that include the $q^2$ dependence~\cite{Bacchio:2025auj}. Specifically, $|V_{us}|$ is extracted using their Eq.~(A20). 
To remain consistent with the lattice calculation, we apply the corresponding electroweak radiative correction via $G_F\equiv G_F\cdot(1+C)\equiv G_F\cdot(1+0.0105)$~\cite{Garcia:1985xz,Bacchio:2025auj}. 


{\setlength{\parskip}{15pt}
\noindent
\textbf{Systematic uncertainties of branching fraction determination}\\ 
Systematic studies have been performed separately for the branching fraction and form factors. The corresponding results are presented in Tables~\ref{tab:sys_bf}, \ref{tab:sys_ff} and \ref{tab:sys_ff2}. For the branching fraction, the uncertainties arise from the requirement on the number of tracks and $\Lambda$ vertex fit, the proton and electron detection,
the 4C-fit, and the estimation of the number of candidates using ST and DT techniques.
\\
{\bf 1. 
Requirement on the number of tracks and $\bm{\Lambda}$ vertex fit.
} The control sample of $J/\psi\to\Lambda(\to p\pi^-)\bar{\Lambda}(\to\bar{p}\pi^+)$ is used to study the systematic uncertainty due to the requirement of four final tracks and the $\Lambda$ reconstruction through the vertex fit. For the requirement on the number of tracks, the MC efficiency is corrected by multiplying the correction factor obtained from the control sample, while the uncertainty from the correction factor is assigned as the systematic uncertainty. The systematic uncertainty due to the $\Lambda$ vertex fit is assigned to be the efficiency difference between the data and MC.\\
{\bf 2. $\bm{p}$ and $\bm{e^{-}}$ detection.
} 
To evaluate the systematic uncertainty from the track reconstruction of the protons and electrons, and their corresponding anti-particles, we use the two control channels $J/\psi\to p\bar{p}\pi^-\pi^+$ and $e^+e^-\to e^+e^-\gamma$. From these, two-dimensional efficiency plots ($\cos\theta_{\mathrm{LAB}}$,$p_\mathrm{T}$), one for each particle species is obtained. For each track, the efficiency plots are used to obtain a correction factor. 
For the electron, the correction factor is sizable, i.e., ($6.81\pm1.65$)\%; we therefore take the corrected result as the nominal value and assign the magnitude of the correction as the systematic uncertainty. For the proton, the correction is small, so the uncorrected result is retained as the nominal value, and the corresponding correction-induced shift is included in the systematic uncertainty.
\\
{\bf 3. Kinematic fit.} 
To test the better agreement of kinematic variables between data and MC samples, the simulated sample is corrected using the correction parameters for the charged tracks following the procedure from~\cite{BESIII:2012mpj}. The systematic uncertainty is assigned to be the difference between the fit results with and without the application of correction parameters.\\
{\bf 4. ST and DT yield extraction.} 
The extracted single-tag and double-tag event yields depend on the fitting functions used to model the kinematically constrained invariant mass and $U_{\rm miss}$ spectra. To evaluate the associated systematic uncertainties, we vary the signal and background line shapes in these fits. For the signal component in the $U_{\rm miss}$ distribution, the Gaussian convolved with the simulated line shape is removed, and only the simulated shape is used, resulting in a 0.10\% change in the yield. An analogous procedure is applied to the background: first, the linear component is omitted, leaving only the simulated non-leptonic line shape, and as an additional test, the background is modelled with a third-order Chebyshev polynomial, which produces a larger deviation of 0.79\%. The total systematic uncertainty associated with the $U_{\rm miss}$ fit is therefore taken to be 0.80\%. For the kinematically constrained invariant mass fit, the signal shape uncertainty is estimated by varying the width of the convolved Gaussian, yielding a 0.01\% change in the branching fraction. Replacing the background model with a simulated Monte Carlo shape results in a 0.37\% variation. The total systematic uncertainty from this fit is assigned to be 0.37\%. To test for potential bias in the fitting procedure, 300 simulated signal-like samples with the same event yields as observed in data are generated and fitted. The resulting fit parameters are consistent with the generated values, and no indication of systematic bias is observed.\\

{\setlength{\parskip}{15pt}
\noindent
\textbf{Systematic uncertainties of form factor determination}\\
Systematic uncertainties are assigned based on the electron momentum, parametrization of the $q^2$ dependence, the estimator, and the fit method. We also performed cross checks on the kinematic fit and decay length, but found no significant systematic effect.\\
{\bf 1. Consistency check.} For the form-factor measurement, 400 pseudo-data samples incorporating the full production and decay dynamics are generated and analysed using the maximum log-likelihood method. The distributions of the fitted parameter values are described by Gaussian functions, one for each form factor. The differences between the Gaussian mean values and the corresponding generated input parameters are assigned as systematic uncertainties.
\\
{\bf 2. Functional dependence of $\bm{q^2}$.} 
The form factors depend on $q^2$, and our nominal result uses a dipole parametrization. We test two alternative descriptions based on the "z-expansion". In the first, the expansion parameters are taken from a relativistic quark model based on a quasipotential approach with a QCD-motivated potential~\cite{Faustov:2018dkn}. In the second, they are taken from the lattice QCD calculation of Ref.~\cite{Bacchio:2025auj}. The larger deviation from the nominal result is assigned as the systematic uncertainty.\\
{\bf 3. Fit method.} The fit method introduces a systematic uncertainty from three different sources: fixed value of production and $\alpha_+(\alpha_-)$, parameters, number of background candidates from $\Lambda\to p \pi^-$ and other sources, and MC efficiency correction factors. 
The contribution due to the uncertainty from these three sources is obtained by repeating the fit procedure 100 times after randomly varying all parameters within their uncertainties.
A Gaussian function is used to fit each form factor and its width is taken as the systematic uncertainty.\\
{\bf 4. $\bm{e^{+}}$ and $\bm{e^{-}}$ reconstruction.} 
To study the performance of electron and positron reconstruction, we vary the momentum selection criteria for the $e^\pm$ candidates. The stability of the result is first tested by scanning a $\pm$25~${\rm MeV}/c$ window around the nominal value of 100~${\rm MeV}/c$. Given the momentum resolution of approximately 2~${\rm MeV}/c$, the requirement is then tightened to a $\pm$5~${\rm MeV}/c$ window. The systematic uncertainty is assigned as the largest deviation from the nominal result.
\\

\begin{table*}[htp]
  \begin{center}
    \caption{\textbf{Systematic uncertainty for the branching fraction measurement, relative to the central value, and the sum in quadrature.}} 
    \renewcommand\arraystretch{1}
    \setlength{\tabcolsep}{6mm}{
      \begin{tabular}{cc}
        \hline\hline
          Source    &		Uncertainty(\%)\\
        \hline
        Number of tracks&			$0.03$ \\
        $\Lambda$ vertex fit&		$0.20$\\
        $p$ tracking&	$0.26$\\
        $e$ detection&	$1.55$\\
        Kinematic fit&	$0.22$\\
        ST fit method&	$0.37$\\
        DT fit method&	$0.80$\\
        \hline
        Sum syst.&	$1.83$ \\
        \hline\hline
      \end{tabular}
    }\label{tab:sys_bf}
  \end{center}
\end{table*}

\begin{table*}[htp]
  \begin{center}
    \caption{\textbf{Absolute systematic uncertainties contribution to the first set of form factor ratios, and the final sum, where the individual contributions are taken in quadrature. The first row shows the statistical uncertainty for reference.}} 
    \renewcommand\arraystretch{1.2}
    \setlength{\tabcolsep}{2.2mm}{
      \begin{tabular}{ccccccc}
        \hline\hline
         & $g^-_{av}$&$g^-_w$& $g^+_{av}$&$g^+_w$& $\langle g_{av}\rangle$&	 $\langle g_w\rangle$\\
         \hline
         Statistical & $^{+0.075}_{-0.057}$& $^{+0.510}_{-0.506}$ & $^{+0.073}_{-0.069}$& $^{+0.496}_{-0.486}$ & $^{+0.048}_{-0.047}$& $^{+0.356}_{-0.353}$\\
        \hline
        Consistency check & $0.001$& $0.007$ & $0.004$& $0.003$ & $0.001$& $0.017$\\
        $q^2$ parametrization & $0.001$& $0.011$ & $0.002$& $0.001$ & $0.001$& $0.004$\\
        Fit method & $0.004$& $0.148$ & $0.012$& $0.183$ & $0.006$& $0.118$\\
        $p_e$ selection & $0.008$&	$0.072$ & $0.006$&	$0.087$ & $0.004$&	$0.067$\\
        \hline
        Sum syst.&	$0.009$&	$0.165$ &	$0.014$&	$0.203$ &	$0.007$&	$0.137$\\
        \hline\hline
      \end{tabular}
    }\label{tab:sys_ff}
  \end{center}
\end{table*}

\begin{table*}[htp]
  \begin{center}
    \caption{\textbf{Absolute systematic uncertainties contribution to the second set of form factor ratios, and the final sum, where the individual contributions are taken in quadrature. The first row shows the statistical uncertainty for reference.}} 
    \renewcommand\arraystretch{1.2}
    \setlength{\tabcolsep}{6mm}{
      \begin{tabular}{cccc}
        \hline\hline
         & $\langle g_{av}\rangle$&	 $\langle g_w\rangle$& $\langle g_{av2}\rangle$\\
         \hline
         Statistical &  $^{+0.089}_{-0.086}$ & $^{+0.529}_{-0.492}$& $^{+0.651}_{-0.629}$\\
        \hline
        Consistency check &  $0.007$ & $0.001$& $0.054$\\
        $q^2$ parametrization& 
         $0.025$ & $0.096$& $0.158$\\
        Fit method &  $0.001$ & $0.079$& $0.044$\\
        $p_e$ selection & $0.005$ & $0.070$&	$0.049$\\
        \hline
        Sum syst.&	$0.026$ &	$0.143$&	$0.179$\\
        \hline\hline
      \end{tabular}
    }\label{tab:sys_ff2}
  \end{center}
\end{table*}


%% file: authorlist_2025-05-08.tex
M.~Ablikim$^{1}$, M.~N.~Achasov$^{4,c}$, P.~Adlarson$^{78}$, X.~C.~Ai$^{83}$, R.~Aliberti$^{36}$, A.~Amoroso$^{77A,77C}$, Q.~An$^{74,60,a}$, Y.~Bai$^{58}$, O.~Bakina$^{37}$, Y.~Ban$^{47,h}$
% ...
\\

\par\noindent\hrulefill

\footnotesize\noindent{\it
$^{1}$ Institute of High Energy Physics, Beijing 100049, People's Republic of China\\
% ...
$^{4}$ Budker Institute of Nuclear Physics SB RAS (BINP), Novosibirsk 630090, Russia\\
% ...
$^{36}$ Johannes Gutenberg University of Mainz, Johann-Joachim-Becher-Weg 45, D-55099 Mainz, Germany\\
$^{37}$ Joint Institute for Nuclear Research, 141980 Dubna, Moscow region, Russia\\
% ...
$^{47}$ Peking University, Beijing 100871, People's Republic of China\\
% ...
$^{58}$ Southeast University, Nanjing 211100, People's Republic of China\\
% ...
$^{60}$ State Key Laboratory of Particle Detection and Electronics, Beijing 100049, Hefei 230026, People's Republic of China\\
% ...
$^{74}$ University of Science and Technology of China, Hefei 230026, People's Republic of China\\
% ...
$^{77}$ University of Turin and INFN, (A)University of Turin, I-10125, Turin, Italy; (B)University of Eastern Piedmont, I-15121, Alessandria, Italy; (C)INFN, I-10125, Turin, Italy\\
$^{78}$ Uppsala University, Box 516, SE-75120 Uppsala, Sweden\\
% ...
$^{83}$ Zhengzhou University, Zhengzhou 450001, People's Republic of China\\
$^{a}$ Deceased\\
% ...
$^{c}$ Also at the Novosibirsk State University, Novosibirsk, 630090, Russia\\
% ...
$^{h}$ Also at State Key Laboratory of Nuclear Physics and Technology, Peking University, Beijing 100871, People's Republic of China\\
% ...

}

%% file: acknowledgement_2025-05-08.tex

The authors thank W. Wang for helpful discussions. The BESIII Collaboration thanks the staff of BEPCII (https://cstr.cn/31109.02.BEPC) and the IHEP computing center for their strong support. This work is supported in part by National Key R\&D Program of China under Contracts Nos. 2023YFA1606704, 2023YFA1606000; National Natural Science Foundation of China (NSFC) under Contracts Nos. 12375092, 11635010, 11935015, 11935016, 11935018, 12025502, 12035009, 12035013, 12061131003, 12192260, 12192261, 12192262, 12192263, 12192264, 12192265, 12221005, 12225509, 12235017, 12361141819; Sichuan Science and Technology Program under Contract No. 2026NSFSC0755; Research Foundation of Southwest University of Science and Technology under Grant No. 25zx7184; Beijing Natural Science Foundation of China (BNSF) under Contract No. IS23014; The Chinese Academy of Sciences (CAS) Large-Scale Scientific Facility Program; CAS under Contract No. YSBR-101; 100 Talents Program of CAS; The Institute of Nuclear and Particle Physics (INPAC) and Shanghai Key Laboratory for Particle Physics and Cosmology; ERC under Contract No. 758462; German Research Foundation DFG under Contract No. FOR5327; Istituto Nazionale di Fisica Nucleare, Italy; Knut and Alice Wallenberg Foundation under Contracts Nos. 2021.0174, 2021.0299; Ministry of Development of Turkey under Contract No. DPT2006K-120470; National Research Foundation of Korea under Contract No. NRF-2022R1A2C1092335; National Science and Technology fund of Mongolia; Polish National Science Centre under Contract No. 2024/53/B/ST2/00975; STFC (United Kingdom); Swedish Research Council under Contracts Nos. 2019.04595, 2021.04567; the Olle Engkvist Foundation under Contract No. 200-0605; U. S. Department of Energy under Contract No. DE-FG02-05ER41374.